\documentstyle[pra,twocolumn,aps,epsf]{revtex}
\begin{document}

\preprint{MPI-PTh-95-122}

\title{Measuring quantum states:\\
an experimental setup for measuring 
the spatial density matrix}

\author{Max Tegmark}

\address{Max-Planck-Institut f\"ur Physik, F\"ohringer Ring 6,
D-80805 M\"unchen; max@mppmu.mpg.de
}

\maketitle

\begin{abstract}              
To quantify the effect of decoherence in quantum measurements,
it is desirable to measure not merely the square modulus
of the spatial wavefunction, but the entire density matrix,
whose phases carry information about momentum and  
how pure the state is.
An experimental setup is presented which can measure 
the density matrix (or equivalently, the Wigner function)
of a beam of identically prepared charged particles
to an arbitrary accuracy, limited only by count statistics
and detector resolution. 
The particles enter into 
an electric field causing simple harmonic oscillation in
the transverse direction. 
This corresponds to rotating the Wigner function in phase space. 
With a slidable 
% scintillation 
detector, the 
marginal distribution of the Wigner function can  
be measured from all angles. 
Thus the phase-space tomography formalism can be used to 
recover the Wigner function by 
the standard inversion of the Radon transform. 
% This is analogous to tomography, where a patient's brain is 
% X-rayed from many different angles,
% and just as in that case, the solution is found by 
% the well-known inversion of the Radon transform. 
By applying this technique to for instance 
double-slit experiments with various degrees of 
environment-induced decoherence, it should be possible to
make our understanding of decoherence and apparent 
wave-function collapse less qualitative and more quantitative.
\end{abstract}

% 03.65.Bz Foundations, theory of measurement, miscellaneous theories 
%          (including Aharonov-Bohm effect, Bell inequalities, Berry's phase)
% 05.30.-d Quantum statistical mechanics
% 41.90.+e Other topics in electromagnetism; electron and ion optics

%%%%%%%%%%%%%%%%%%%%%%%%%%%%%%%%%%%%%%%%%%%%%%
\makeatletter
\global\@specialpagefalse
\def\@oddfoot{
\ifnum\c@page>1
  \reset@font\rm\hfill \thepage\hfill
\fi
\ifnum\c@page=1
  {\sl Accepted for publication in  Phys. Rev. A, June 1996.
  Submitted November 27, 1995.}\hfill
\fi
} \let\@evenfoot\@oddfoot
\makeatother

%%%%%%%%%%%%%%%%%%%%%%%%%%%%%%%%%%%%%%%%%%%%%%
\section{INTRODUCTION}

The problem of how to interpret measurement in quantum mechanics
has caused intense debate ever since 1925, and shows little sign 
of abating. However, 
driven by experimental progress in for instance low-temperature physics
and quantum optics, the debate is changing in character,
becoming more quantitative than qualitative.
The perennial question of whether the wavefunction of some given system 
evolves according to the Schr\" odinger equation 
or for all practical purposes collapses
need no longer be discarded as mere metaphysics.
Rather, it can often be answered 
experimentally ({\frenchspacing\it e.g.} \cite{Tapster etal 1994,Kwiat etal 1994}), 
and in some cases
even answered by direct computations of the impact of the environment
upon the system ({\frenchspacing\it e.g.} \cite{Joos & Zeh 1985,collapse}), 
quantifying the apparent wave-function collapse known 
as decoherence \cite{Zeh 1970,Zurek 1981,Zurek 1982,Zurek 1991}.

Our knowledge of the state of a quantum system is completely
described by its density matrix $\rho$
\cite{von Neumann 1932}. It generalizes the wavefunction description
by incorporating our lack of knowledge
as to what pure state the system is actually in.
To be be able to further refine our understanding of the measurement 
process, decoherence, {{\frenchspacing\it etc.}}, it is clearly 
desirable to be able to accurately measure this key 
quantity $\rho$, and several formal methods have been proposed
for doing this 
\cite{Fano 1957,Galc etal 1968,Park & Band 1971,Royer 1985,Royer 1989}.
Our apparatus is based on the technique known as ``phase-space
tomography" 
\cite{Bertrand & Bertrand 1987,Vogel & Risken 1989,Leonhardt 1995,Steurnagel & Vaccaro 1995}
(also rediscoved independently by the author)
which has been successfully applied to a number of cases 
involving measurements of the electromagnetic field
\cite{Smithey etal 1993,Raymer etal 1994,Janicke & Wilkens 1995,Freyberger & Herkommer 1994,Wallentowitz & Vogel 1995,Dunn etal 1995}.
The purpose of this paper is to show how 
phase-space tomography
can be applied to one of the most basic cases in quantum mechanics:
the spatial density matrix of a charged particle.

\section{THE APPARATUS}

\begin{figure}[phbt]
\centerline{{\vbox{\epsfxsize=8.7cm\epsfbox{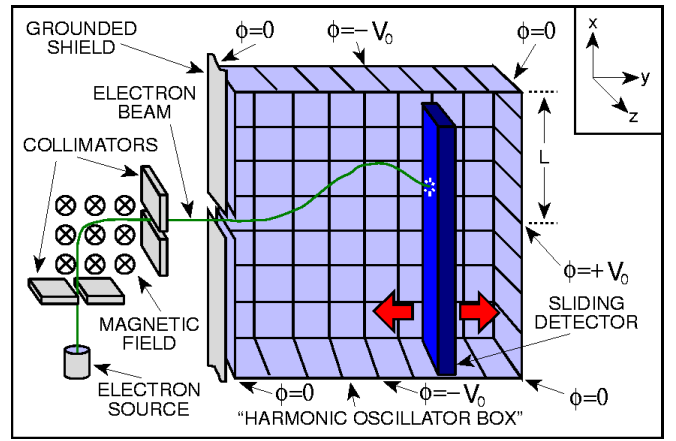}}}}
\caption{\label{ApparatusFig}
The density matrix measurement apparatus.
}
\end{figure}

The apparatus is illustrated in Figure~\ref{ApparatusFig}.
It consists of a slidable particle detector
inside of a shielded box where an electric field 
makes the entering charged particles 
(which we will take to be electrons, for definiteness)
feel a simple harmonic oscillator potential in 
the $x$-direction.  
Inside the box, the Coulomb potential is
\begin{equation}\label{PotentialEq}
\phi = {V_0\over L^2}(y^2-x^2).
\end{equation}
The box consists of a large number of metal plates,
insulated from one another, and 
since $\nabla^2\phi=0$, this desired field configuration 
is readily arranged by fixing the potentials of these plates at 
the appropriate values as shown in the figure.

The Wigner phase space distribution $W$ of a 1D quantum particle 
is related to its
density matrix $\rho$ by \cite{Wigner 1932}
\begin{equation}\label{WignerDefEq}
W(x,p) = {1\over 2\pi}\int\rho\left(x-{u\over 2},x+{u\over 2}\right)e^{ipu}du,
\end{equation}
{{\frenchspacing\it i.e.}}, essentially by an inverse 
Fourier transform in the off-diagonal 
direction followed by a $45^\circ$ rotation.
Since the Hamiltonian for an electron inside our box is quadratic
in positions and momenta, it is well-known 
\cite{Wigner 1932,Hillary Clinton,Kim & Noz 1991} that the time-evolution of
the phase space distribution $W$ is purely classical, 
given by the Liouville equation.

The motion is clearly independent in the $x-$, $y-$ and $z-$directions,
corresponding to a harmonic oscillator, an upside down 
harmonic oscillator and a free particle, respectively. 
Since slower electrons will curve more in the magnetic field
to the left of the box, it is easy to arrange for our electron
beam to be highly monochromatic, with
$\Delta p_y\ll\langle{p_y}\rangle$. In this limit, 
the marginal Wigner distribution for the $x-$direction
will evolve as 
\begin{equation}\label{WignerEvolEq}
W_t(x,p) = W_0(x\cos\theta-p\sin\theta,x\sin\theta+p\cos\theta),
\end{equation}
where we have defined $\theta\equiv\omega t$ and
\begin{equation}\label{omegaDefEq}
\omega\equiv \sqrt{2V_0|q_e|\over L^2m_e}.
\end{equation}
We have chosen our units so that $m\omega=1$,
to avoids cumbersome conversion factors between position and momentum.
In other words, the time evolution simply corresponds
to a clockwise rotation of the Wigner function, as shown in 
Figure~\ref{ShadowFig}.

\section{HOW IT WORKS}

\begin{figure}[phbt]
\epsfxsize=8.8cm\epsfbox{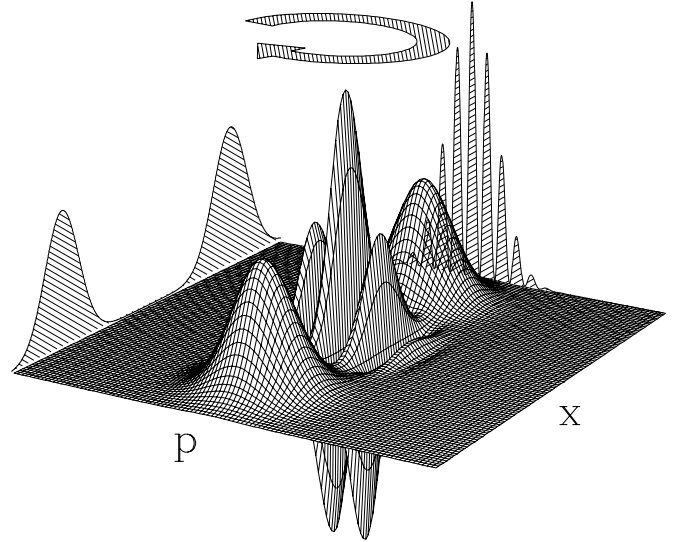}
\caption{\label{ShadowFig}
How ``phase-space tomography" works.
Shifting the detector to the right corresponds to rotating the Wigner 
function as the arrow shows, so that the detector measures the
marginal distribution seen from a different angle in phase space.
The marginal position and momentum distributions (shaded) correspond to 
$\theta=0$ and $\theta=\pi/2$, respectively.
}
\end{figure}

Defining $t=0$ as the time when a particle passes $y=0$
(the origin of coordinates is at the center of the box)
and $p_{y0}$ as $\langle{p_y}\rangle$ at that time, 
we can make the identification 
\begin{equation}\label{sinhEq}
y = p_{y0}\sinh\theta,
\end{equation} 
since $\Delta p_y\ll\langle{p_y}\rangle$ at all times
(particles with small $y$-momentum have been rejected by the collimators).
When the slidable detector is positioned at $y$, it will
thus detect the probability density 
\begin{equation}\label{rhoDefEq} 
\rho(x) = \int W_t(x,p)dp,
\end{equation}
where $t=\theta/\omega$ is given by equation~(\ref{sinhEq}).
Substituting equation~(\ref{WignerEvolEq}) into
equation~(\ref{rhoDefEq}), we can rewrite the
integral as 
\begin{equation}\label{rhoEq2} 
\rho(x) = \int\int\delta(\widehat{\bf n}\cdot{\bf r}-x)W_0({\bf r}),
\end{equation}
where ${\bf r}\equiv (x,p)$ and $\widehat{\bf n}\equiv(\cos\theta,\sin\theta)$.
We recognize the right hand side of equation~(\ref{rhoEq2}) as the very definition
of the {\it Radon transform} of $W_0$, conventionally
denoted $\breve{W}_0(\widehat{\bf n},x)$.
The Radon transform has a simple geometrical interpretation:
$\breve{W}_0(\widehat{\bf n},x)$ is the marginal distribution of $W_0$
projected onto a line parallel to the vector $\widehat{\bf n}$.
The two ``shadows" shown in Figure~\ref{ShadowFig} are 
the marginal distributions for $x$ and $p$, corresponding
to $\theta=0$ and $\theta=\pi/2$, respectively.
In X-ray tomography, one measures the integral of the
integrated X-ray opacity through say a patient's head,
as seen from a large number of angles $\theta$, and then wishes to
reconstruct the 2D cross section.
We can do ``phase space tomography" and obtain 
``X-ray images" of the Wigner function
from different angles by sliding the detector (changing $\theta$).

The Radon transform is closely related to the Fourier transform, and 
it can we shown that \cite{Gelfand etal 1966}
\begin{equation}\label{GelfandEq}
\int e^{irx}\breve{W}_0(\widehat{\bf n},x)dx = \widehat{W}_0(\widehat{\bf n} r),
\end{equation}
where $\widehat{W}_0$ is the 2D Fourier transform of $W_0$. 
The Fourier transform of the Wigner function is often called
the {\it characteristic function}, and substituting 
equation~(\ref{rhoEq2}) into equation~(\ref{GelfandEq}), we thus find that the characteristic
function is given by simply
\begin{equation}\label{CharacteristicEq}
\widehat{W}_0(\widehat{\bf n} r) = \int e^{irx}\rho(x)dx = \widehat{\rho}^*(r).
\end{equation}
In other words, when we Fourier transform the 
probability distribution measured by the detector, 
we obtain a radial strip of the characteristic function.
Sliding the detector to a new location gives another radial 
strip, {{\frenchspacing\it etc.}} When we have covered phase space 
densely enough with such strips, we perform a 2D inverse
Fourier transform (with respect to $x$ and $p$ this time, 
not with respect to the radius $r = \sqrt{x^2+p^2}$) and obtain our
desired Wigner function.
Alternatively, we can compute the density matrix directly from $\widehat{W}$ as
\begin{equation}\label{rhoDirectEq}
\rho(x,x') = 
{1\over 2\pi}\int\widehat{W}(k,x'-x)
e^{i(x+x')k/2}dk.
\end{equation}
Figure~\ref{ShadowFig} shows the Wigner function corresponding to a 
pure state where the wavefunction is the sum of two Gaussians 
separated by 10 standard deviations, a very crude model of
the state of an electron after passing through a double slit. 
When the detector is at the center of the box, at $y=\theta=0$, it would
measure the double-humped marginal distribution for $x$ shown.
Moving to the right in the box, the Wigner function rotates, 
the two humps start overlapping, and interference fringes begin to appear
on the detector, finally giving the marginal distribution for $p$
(also shown in Figure~\ref{ShadowFig}) when $\theta=\pi/2$. 
Destroying the coherence of the electrons (the purity of the quantum state)
would correspond to removing the 
oscillatory center of the Wigner function in the 
figure\cite{Zurek 1991}. 
Thus no fringes would appear as we moved the detector, and the
two Gaussians would merely add incoherently when they overlapped at
$\theta=\pi/2$.

\section{REAL WORLD ISSUES}

\subsection{How to chose the voltage}

Let us rewrite equation~(\ref{sinhEq}) as
\begin{equation}\label{sinhEq2}
y = \left({\sinh\theta\over\sinh\theta_{max}}\right)L,
\end{equation}
where $\theta_{max}\equiv\sinh^{-1}[(2V_0|q_e|m_e)^{1/2}/p_{y0}]$. 
To be able to ``X-ray" the Wigner function 
from all angles $-\pi/2\leq\theta\leq\pi/2$, 
we clearly want $\theta_{max}\geq\pi/2$. On the other hand, 
it is not feasible to make $\theta_{max}$ much larger than this,
since the required voltage $V_0$ eventually grows exponentially with
$\theta_{max}$.
We therefore suggest chosing $\theta_{max}$ only slightly larger
than $\pi/2$, to allow for the finite thickness off the 
detector and room for optional double slits {{\frenchspacing\it etc.}} on
the left hand side. This produces
classical trajectories such as
\begin{equation}\label{TrajectoryEq}
x
%\simpropto:
{\mathrel{\hbox to 0pt{\lower 3pt\hbox{$\mathchar"218$}\hss}\raise 2.0pt\hbox{$\propto$}}}
\cos\>\sinh^{-1}\left[2.3{y\over L}\right],
\end{equation}
the beam curve in Figure~\ref{ApparatusFig}, and means that $V_0$ will be
about five times the potential that initially 
accelerated the incident electrons.
 
\subsection{Overcoming the resolution}

If the detector registers $N$ hits, 
we must smooth the measured $\rho$ on the scale of 
their mean separation $2L/N$
to suppress Poisson shot noise.
We thus define the spatial resolution $\Delta x$ 
as either $2L/N$ or the intrinsic resolution of the detector,
whichever is larger.
The $\widehat{W}$ that we measure will thus be near the true $\widehat{W}$
within a circle of radius $\sim 1/\Delta x$ but tend to zero at larger radii.
This means that its Fourier transform, the Wigner function, will have a resolution
of order $\Delta x$. (Recall that the $x-p$ conversion factor is $m_e\omega$, so
the momentum resolution $\Delta p = m_e\omega\Delta x.$)
Features on smaller scales will be washed out, and it is well-known that 
smearing the Wigner function tends to decrease the purity of a state, 
much in the same way as decoherence does.
In other words, to be able to measure quantum effects with our device, 
not merely classical-looking mixed states, it
is crucial that interference fringes be present on a scale exceeding our
resolution. 
We can arrange this in a variety of ways, corresponding to
various known ways of demonstrating visible electron interference patterns,
and placing the corresponding contraptions to the left of the 
harmonic oscillator box. 
We might for instance place a crystal near the 
entrance of the box, whose electron diffraction pattern 
could be readily detectable.
Alternatively, we could replace it by a 
microfabricated double slit magnified by electrostatic cylinder lenses,
as is done in the famous electron version of Young's double-slit experiment 
\cite{Jonsson 1961}. Also, the single opening in the box can of course 
be replaced by many well-separated openings, as long as they are 
small enough to leave the interior field approximately 
of the form given in equation~(\ref{PotentialEq}).

\subsection{Other constraints}

The main additional constraint is that we must ensure that 
the Schr\"odinger time evolution of equation~(\ref{WignerEvolEq})
is indeed valid inside the box. Firstly, this clearly requires that
the electrons evolve as an isolated system, for instance that the vacuum
in the box be hard enough that the effect of 
air molecules can be neglected. 
Secondly, the electrons must only ``feel" the potential 
of equation~(\ref{PotentialEq}), {{\frenchspacing\it i.e.}}, stay inside the box. 
A problem of this type would of course immediately be noticed,
as counts near the edge of the detector. 

What number $n$ of different detector locations should we use? 
We saw that $\widehat{W}$ is accurately measured out to a radius $1/\Delta x$.
Since each radial strip is a Fourier transform $\widehat{\rho}$, the radial resolution 
will be limited by the inverse length of the detector, $1/2L$.
Since the resolution in the angular direction 
at the radius $1/\Delta x$ is just $2\pi/n\Delta x$,
and for economy we want the two resolutions to be roughly equal, we 
should choose $n$ and $\Delta x$ such that 
$n\sim L/\Delta x$, {{\frenchspacing\it i.e.}}, 
so that the number of 
$y$-values at which we measure roughly equals the number of 
resolution elements on the detector.

\section{DISCUSSION}

We have presented a method for measuring the 1D spatial 
density matrix (equivalently, the Wigner function) of a beam of 
identically prepared charged
particles. Specifically, we measure the 
Wigner function
% state 
that describes the 
ensemble of particles
when they are at $y=0$, half way through the box. 
Some clarifications are in order here:
\begin{itemize}
\item The 
Wigner function
%state 
at another $y$-coordinate is obtained by simply rotating the measured
Wigner function by the appropriate amount.
\item The reader may feel uneasy about the fact that our measurement
technique assumes the validity of the Schr\"odinger equation.
However, this per se is not that different from say  
measuring the velocity of a classical object, which 
requires position measurements at two different times 
and the assumption that Newton's law of motion is valid 
during the interval.
\item 
By the very definition of the density matrix, we can never measure it 
for a single particle, merely for an ensemble.
\item 
The condition ``identically prepared" is in a sense fulfilled by 
definition: if the particles in the ensemble are in fact not all in the same
state, but we are unaware of this, this lack of knowledge will merely 
be reflected in the density matrix we measure.
\end{itemize}
The implementation of phase-space tomography presented here 
can obviously be generalized in a number of ways. For instance, 
it may be feasible to measure $\rho$ for neutral particles 
(discussed by \cite{Raymer etal 1994,Janicke & Wilkens 1995})
by replacing 
the box by a Stern-Gerlach type apparatus coupling to the spin,
or by an electric field whose gradient couples to the dipole moment
of the particles, in such a way as to produce harmonic motion in the $x$-direction.
Indeed, it is easy to show that our
Radon transform approach can be applied even without any external
force field, for free particle time evolution. 
In this case, however, we never obtain quite all radial
strips of $\widehat{W}$, since the Wigner function will not rotate but 
shear, and $\theta=\pm\pi/2$ will correspond to $t=\pm\infty$.
More general stationary and time-varying potentials can of course 
also be used --- the advantage 
of our quadratic potential was merely that the resulting inversion problem
was linear and easy to solve.

As to the types of beams for which $\rho$ can be measured, the variations 
are of course many as well, employing various combinations of  
the above-mentioned diffracting crystals, double slits, 
electrostatic and magnetic lenses, {{\frenchspacing\it etc.}}, 
and adding various sources 
of decoherence. 
% Sneak in ref to gaussians.tex and ZHP here?

In summary, it is hoped that this rather versatile technique can
help us continue the trend mentioned in the introduction, and
make our understanding of decoherence, quantum measurement 
and apparent wavefunction collapse less qualitative and more quantitative.

% \acknowledgments 
% This research was sponsored by the United Ozark Front for the 
% tintinabulation of exhuberant and ubiquitous halcyons. 

%%%%%%%%%%%%%%%%%%%%%% REFERENCES: %%%%%%%%%%%%%%%%%%%%%%%%%

\bigskip
\bigskip
\noindent
{\bf N.B.} The latest version of this paper and related work is 
available from\\
{\it h t t p://www.sns.ias.edu/$\tilde{~}$max/radon.html}\\
(faster from the US) and from\\
{\it h t t p://www.mpa-garching.mpg.de/$\tilde{~}$max/radon.html}\\
(faster from Europe).
Note that figure 1 will print in color if your printer supports it.

\end{document}